\documentstyle{article}
\begin{document}
\title{The {C}hurch-{T}uring thesis as a guiding principle for physics}
\author{Karl Svozil\\
Institut f\"ur Theoretische Physik  \\
University of Technology Vienna      \\
Wiedner Hauptstra\ss e 8-10/136    \\
A-1040 Vienna, Austria              \\
e-mail: svozil@tph.tuwien.ac.at\\
www: http://tph.tuwien.ac.at/$\widetilde{\;}$svozil}
\date{ }
\maketitle

\begin{flushright}
{\scriptsize
http://tph.tuwien.ac.at/$\widetilde{\;}$svozil/publ/ct.tex}
\end{flushright}

\begin{abstract}
Two aspects of the physical side of the Church-Turing
thesis are discussed.
The first issue is a variant of the Eleatic argument against motion,
dealing with Zeno squeezed time cycles of computers.
The second argument reviews the issue of one-to-one computation, that
is, the bijective (unique and reversible) evolution of computations and
its relation to the measurement process.
\end{abstract}

\section{Introduction}
It is reasonable to require
from a ``useful'' theory of computation that any
capacity and feature of physical systems (interpretable as ``computing
machines'') should be reflected therein and
{\it vice versa.}

The recognition of the physical aspect of the Church-Turing thesis---the
postulated equivalence between the informal notion of ``mechanical
computation'' (algorithm) and recursive function theory as its formalized
counterpart---is not new \cite{rogers1,odi:89}.
In particular Landauer,
inspired by the books of
Brillouin
\cite{brillouin1,brillouin2}
points out that computers are physical systems, that computations are
physical processes and therefore are subject to the laws of physics
\cite{landauer:61,landauer-67,landauer:82,landauer-87,landauer-88,landauer-89,landauer,landauer-94,landauer-95}.
 As Deutsch puts it \cite[p. 101]{deutsch},
 \begin{quote}
 {\em
 ``The reason why we find it possible to construct, say, electronic
 calculators, and indeed why we can perform mental arithmetic, cannot
 be found in mathematics or logic. {\em
 The reason is that the laws of physics `happen to' permit the
 existence of physical models for the operations of arithmetic}
 such as addition, subtraction and multiplication.
 If they did not, these familiar operations would be
 noncomputable functions. We might still
 know {\em of} them and invoke them in mathematical proofs
 (which would presumably be called `nonconstructive') but we could
 not perform them.''
 }
 \end{quote}
See also Pitowsky's review \cite{pit:90}.

The Computer Sciences are well aware of this
connection. See, for instance,
Odifreddi's review
\cite{odi:89}, the articles by
 Rosen \cite{rosen} and Kreisel \cite{kreisel},
 or
 Davis'
 book \cite[p. 11]{davis-58}, where the following question is
asked:
 \begin{quote}
 {\em `` $\ldots$ how can we ever exclude the possibility of our
 presented,
 some day (perhaps by some extraterrestrial visitors), with a (perhaps
 extremely complex) device or ``oracle'' that ``computes'' a
 noncomputable function?''
 }
 \end{quote}

Thus, it comes as no surprise that the Church-Turing thesis is under
permanent attack from the physical sciences. For just two such attempts
in the recent literature, we refer to the articles by Siegelmann
\cite{siegel95} and Hogarth \cite{hogarth1,hogarth2}.

Even more so, this applies  to the weak Church-Turing thesis, often
referred to as ``Cook-Karp thesis,'' putting into question the
robustness of the notion of tractability or polynomial time complexity
class
with respect to variations of ``reasonable'' models of computation.
One particular famous contemporary case is quantum computing.
There, it has been shown that at least factoring may require polynomial
time on quantum computers within
``reasonable statistics'' \cite{shor:94,ekerj96}.

We shall take out two examples of connections between physics and
computation. First, we briefly review reformulations of Zeno's argument
of Achilles
and the Tortoise (Hector). This paradox purportedly seems to have been
originally directed against motion \cite{zeno,ki-57,gruenbaum:68}.
In this context it can be applied against the uncritical
use of continua and dense sets in general.
Later on, we
shall investigate reversible
computations, more specifically computations corresponding to
bijective (one-to-one)
maps and its possible connections with measurement operations.

\section{Infinity machines by Zeno squeezed time cycles}

In what follows, an
{\em oracle} problem
 solver will be introduced
whose capacity exceeds  and outperforms any presently
 realisable, finite machine and also  any
 universal computer such as the Turing machine. We follow
previous discussions
(cf. \cite[pp. 24-27]{svozil-93} and
\cite{svozil-nat-acad,svozil-complexity:95,svozil-set,maryland}).

 Its
 design is based upon
 a universal computer with ``squeezed'' cycle times of
 computation according to a geometric progression.
 The only difference
between universal computation
 and this type of
 oracle computation is the speed of execution. But what a difference
indeed: Zeno squeezed oracle
 computation performs
 computations {\em in the limit of infinite time of computation}.
 In order to achieve this limit,
 two time scales are introduced: the {\em intrinsic time scale
 of the process of computation}, which approaches infinity in finite
{\em
 extrinsic or proper time of some outside observer}.

As a consequence, certain tasks which lie beyond the domain of recursive
function theory  become computable and even tractable. For example, the
halting problem
and any problem codable into a halting problem would become solvable.
It would also be possible to produce an otherwise
 uncomputable and random output---equivalent to the tossing of a fair
 coin---such as Chaitin's $\Omega$ \cite{chaitin:92,calude:94}
 in finite proper time. We shall come back to these issues, in
particular consistency, shortly.

 A very similar setup has been introduced by Hermann Weyl
 \cite{weyl:49}, which was discussed
 by
 Gr\"unbaum
\cite[p. 630]{gruenbaum:74}.
Already Weyl raised the question whether it is kinematically feasible
for a machine to carry out an {\em infinite} sequence of operations in
{\em finite} time.
Weyl writes \cite[p. 42]{weyl:49},
\begin{quote}
Yet, if the segment of length 1 really consists of infinitely many
sub-segments of length 1/2, 1/4, 1/8,
$\ldots$, as of `chopped-off' wholes, then it is incompatible with the
character of the infinite as the `incompletable' that Achilles should
have been able to traverse them all. If one admits this possibility,
then there is no reason why a machine should not be capable of
completing an infinite sequence of distinct acts of decision within a
finite amount of time; say, by supplying the first result after 1/2
minute, the second after another 1/4 minute, the third 1/8 minute later
than the second, etc. In this way it would be possible, provided the
receptive power of the brain would function similarly, to achieve a
traversal of all natural numbers and thereby a sure yes-or-no decision
regarding any existential question about natural numbers!
\end{quote}

See also
the articles by Thomson \cite{thom:54}, Benacerraf \cite{benna:62},
Rucker \cite{rucker},
Pitowsky \cite{pit:90}, Earman and Norton \cite{ear-nor:93} and Hogarth
\cite{hogarth1,hogarth2},
as well as
 E. W. Beth,
 \cite[p. 492]{beth-59}
 and
K. L\'opez-Escobar
 \cite{le-91}.

 Let us come back to the original goal: the construction of a
``Zeno squeezed oracle,''
 or, in Gr\"unbaum's terminology, of an  ``infinity machine.''
As sketched before, it
can be conceived by considering two
time scales $\tau$ and $t$
which are related as follows.
\begin{description}
\item[$\bullet$]
The
 {\em proper time} $\tau$ measures the physical system time by clocks
 in a way similar to the usual operationalisations; whereas
\item[$\bullet$]
 a discrete {\it cycle time} $t=0,1,2,3,\ldots$ characterizes a sort of
 ``intrinsic'' time scale for a process  running on an otherwise
universal machine.
\item[$\bullet$]
 For some  unspecified
 reason we assume that this machine would allow us to
 ``squeeze'' its intrinsic time  $t$  with respect to the proper time
 $\tau$ by a geometric progression. Hence, for $k<1$, let
 any time cycle of $t$, if measured in terms of $\tau$, be squeezed by
 a factor of $k$ with respect to the foregoing time cycle
i.e.,
 \begin{eqnarray}
 \tau_0&=&0,\quad \tau_1=k ,\quad \tau_{t+1}-\tau_t =k(\tau_t
-\tau_{t-1}),
\\
 \tau_{t}&=&\sum_{n=0}^tk^n-1={k(k^t-1)\over k-1}\quad .
 \end{eqnarray}
Thus, in the limit
 of infinite cycle time $t\rightarrow \infty$, the proper time
 $\tau_\infty = k/(1-k)$ remains finite.
\end{description}
We just mention that for the model introduced here
only
dense space-time would be required.

There is no commonly accepted principle which would forbid
such an oracle {\it a priori}. In particular, classical mechanics
postulates space and time continua as a foundational principle. One
might argue that such an oracle would require a geometric
energy increase  resulting in an infinite consumption of energy. Yet, no
currently accepted physical principle
excludes us from assuming that every geometric
decrease
in cycle time could be associated with a geometric progression in energy
consumption, at least up to some limiting (e.g., Planck) scale.

Nevertheless,
it can be shown by a
diagonalization argument that the
application of
such oracle subroutines
would result in a paradox.
The paradox is constructed in the context of the halting problem.
It is formed in a similar manner as Cantor's diagonalization
argument.
Consider an arbitrary algorithm $B(x)$ whose input is a string of
symbols
$x$.
 Assume that there exists (wrong) an ``effective halting algorithm''
 ${ HALT}$, implementable on the oracle described above,
 which  is
 able to decide whether $B$ terminates on $x$ or not.

 Using  ${ HALT}(B(x))$ we shall construct another
 deterministic computing agent
$A$, which
 has as input any effective program $B$ and which proceeds as follows:
 Upon reading the program $B$ as input, $A$ makes a copy of it.
 This  can be readily achieved, since
 the program $B$ is presented to $A$  in some
 encoded form $\# (B)$, i.e., as a string of symbols. In the next
 step, the agent uses the
 code $\# (B)$ as input string for $B$ itself; i.e., $A$  forms
 $B(\#(B))$, henceforth denoted by $B(B)$. The agent now hands
 $B(B)$ over to its
 subroutine ${ HALT}$.
 Then, $A$ proceeds as follows:
  if ${ HALT}(B(B))$ decides that $B(B)$
 halts, then the agent
 $A$ does not halt;
this can for instance be realized by an infinite {
 DO}-loop;
  if ${ HALT}(B(B))$ decides that $B(B)$
 does {\em not} halt, then
 $A$ halts.

 We shall now confront the agent $A$ with a paradoxical task by
 choosing $A$'s own code as input for itself.---Notice
 that $B$ is arbitrary and has not yet been specified and we are
totally justified to do that:
The deterministic agent $A$ is representable by an algorithm with code
$\# (A)$. Therefore, we are free to substitute $A$ for $B$.

Assume that classically $A$ is restricted to classical bits of
information.
Then, whenever
 $A(A)$ halts,  ${ HALT}(A(A))$  forces
 $A(A)$ not to halt.
Conversely, whenever $A(A)$ does not halt, then ${ HALT}(A(A))$
 steers $A(A)$
 into the halting mode. In both cases one arrives at a
complete contradiction.

Therefore, at least in this example, too powerful physical models (of
computation) are inconsistent.
It almost goes without saying that the concept of infinity
machines is neither constructive nor operational in the current physical
framework.

Quantum mechanics offers a rescue;
yet in a form which is not common in ``classical'' recursion theory.
The paradox is resolved when
$A$ is allowed a nonclassical qubit of information.
Classical information theory is based on the
classical bit as
fundamental atom.
Any classical bit is in one of two
classical states $t$ (often interpreted as ``true'') and $f$ (often
interpreted as ``false'').
In quantum information theory
the most elementary unit of information is
 the {\em quantum bit,}
henceforth called {\em qubit}.
Qubits can be physically represented by a coherent
superposition
of two orthonormal quantum
 states $t$ and $f$.
The quantum bit states
$$
\vert a,b\rangle
=
a\left( \begin{array}{c}1\\ 0\end{array} \right)+
b\left( \begin{array}{c}0\\ 1\end{array} \right)
=
a\vert 0\rangle +b\vert 1 \rangle ,
$$
with
$ \vert a\vert^2+\vert b\vert^2=1$, $a,b\in { C}$
form a continuum.

Assume now that
$\vert 0\rangle =
\vert 1,0\rangle
$
and $\vert 1\rangle =
\vert 0,1\rangle
$
 correspond to the halting and to the nonhalting
states, respectively.
$A$'s task can consistently be performed
if it inputs a qubit corresponding to the {\em fixed point} state
of the diagonalization ($not$) operator
$$
\widehat{D}= not=
\tau_1 =
\left(
\begin{array}{cc}
0 & 1\\
1 & 0
\end{array}
\right) =\vert 1\rangle \langle 0\vert
+ \vert 0\rangle \langle 1\vert    .
$$
That is,
\begin{equation}
\widehat{D}\vert \ast \rangle =\vert \ast \rangle  .
\end{equation}
The fixed point state $\vert \ast \rangle $ is
just the eigenstate of the diagonalization operator
$\widehat{D}$ with
eigenvalue $1$.
Notice that the eigenstates of
$\widehat{D}$ are
\begin{equation}
\vert I\rangle      ,
\vert II\rangle
=
{1\over \sqrt{2}}\left[
\left( \begin{array}{c}1\\ 0\end{array} \right)\pm
\left( \begin{array}{c}0\\ 1\end{array} \right)
\right]
=
{1\over \sqrt{2}}(\vert 0\rangle \pm \vert 1\rangle )
\end{equation}
with the  eigenvalues $+1$ and $-1$, respectively.
Thus, the nonparadoxical, fixed point qubit
in the basis of $\vert 0\rangle $ and $\vert 1\rangle $ is given by
\begin{equation}
\vert \ast \rangle =\vert {1\over \sqrt{2}},{1\over \sqrt{2}} \rangle
=\vert I\rangle .
\end{equation}
In natural language,
this qubit solution corresponds to the statement that
it is impossible
for the agent to control the outcome, since
 there is a
fifty percent chance for each of the  classical bit states $\vert
0\rangle$ and
$\vert 1\rangle$ to be ``stimulated'' at
$t_{A}$.
The impossibility of outcome control
is
indeed encountered
in quantum mechanics.
Stated differently: at the level of probability amplitudes, quantum
theory permits a Zeno squeezed oracle.  But at the level
of observable
probabilities, this is exactly nullified, as
the result of the computation
 appears to
occur entirely at random.

\section{One-to-one computational paths and measurement}

\begin{figure}
\begin{center}
\unitlength 0.70mm
\linethickness{0.4pt}
\begin{picture}(200.00,149.00)
\put(10.00,10.00){\circle*{2.00}}
\put(20.00,10.00){\circle*{2.00}}
\put(30.00,10.00){\circle*{2.00}}
\put(40.00,10.00){\circle*{2.00}}
\put(10.00,20.00){\circle{2.00}}
\put(10.00,11.00){\vector(0,1){8.00}}
\put(10.00,30.00){\circle{2.00}}
\put(10.00,21.00){\vector(0,1){8.00}}
\put(10.00,40.00){\circle{2.00}}
\put(10.00,31.00){\vector(0,1){8.00}}
\put(10.00,50.00){\circle{2.00}}
\put(10.00,41.00){\vector(0,1){8.00}}
\put(10.00,60.00){\circle{2.00}}
\put(10.00,51.00){\vector(0,1){8.00}}
\put(10.00,70.00){\circle{2.00}}
\put(10.00,61.00){\vector(0,1){8.00}}
\put(10.00,80.00){\circle{2.00}}
\put(10.00,71.00){\vector(0,1){8.00}}
\put(10.00,90.00){\circle{2.00}}
\put(10.00,81.00){\vector(0,1){8.00}}
\put(20.00,20.00){\circle{2.00}}
\put(20.00,11.00){\vector(0,1){8.00}}
\put(20.00,30.00){\circle{2.00}}
\put(20.00,21.00){\vector(0,1){8.00}}
\put(20.00,40.00){\circle{2.00}}
\put(20.00,31.00){\vector(0,1){8.00}}
\put(20.00,50.00){\circle{2.00}}
\put(20.00,41.00){\vector(0,1){8.00}}
\put(20.00,60.00){\circle{2.00}}
\put(20.00,51.00){\vector(0,1){8.00}}
\put(20.00,70.00){\circle{2.00}}
\put(20.00,61.00){\vector(0,1){8.00}}
\put(20.00,80.00){\circle{2.00}}
\put(20.00,71.00){\vector(0,1){8.00}}
\put(20.00,90.00){\circle{2.00}}
\put(20.00,81.00){\vector(0,1){8.00}}
\put(30.00,20.00){\circle{2.00}}
\put(30.00,11.00){\vector(0,1){8.00}}
\put(30.00,30.00){\circle{2.00}}
\put(30.00,21.00){\vector(0,1){8.00}}
\put(30.00,40.00){\circle{2.00}}
\put(30.00,31.00){\vector(0,1){8.00}}
\put(30.00,50.00){\circle{2.00}}
\put(30.00,41.00){\vector(0,1){8.00}}
\put(30.00,60.00){\circle{2.00}}
\put(30.00,51.00){\vector(0,1){8.00}}
\put(30.00,70.00){\circle{2.00}}
\put(30.00,61.00){\vector(0,1){8.00}}
\put(30.00,80.00){\circle{2.00}}
\put(30.00,71.00){\vector(0,1){8.00}}
\put(30.00,90.00){\circle{2.00}}
\put(30.00,81.00){\vector(0,1){8.00}}
\put(40.00,20.00){\circle{2.00}}
\put(40.00,11.00){\vector(0,1){8.00}}
\put(40.00,30.00){\circle{2.00}}
\put(40.00,21.00){\vector(0,1){8.00}}
\put(40.00,40.00){\circle{2.00}}
\put(40.00,31.00){\vector(0,1){8.00}}
\put(40.00,50.00){\circle{2.00}}
\put(40.00,41.00){\vector(0,1){8.00}}
\put(40.00,60.00){\circle{2.00}}
\put(40.00,51.00){\vector(0,1){8.00}}
\put(40.00,70.00){\circle{2.00}}
\put(40.00,61.00){\vector(0,1){8.00}}
\put(40.00,80.00){\circle{2.00}}
\put(40.00,71.00){\vector(0,1){8.00}}
\put(40.00,90.00){\circle{2.00}}
\put(40.00,81.00){\vector(0,1){8.00}}
\put(50.00,10.00){\circle*{2.00}}
\put(50.00,20.00){\circle{2.00}}
\put(50.00,11.00){\vector(0,1){8.00}}
\put(50.00,30.00){\circle{2.00}}
\put(50.00,21.00){\vector(0,1){8.00}}
\put(50.00,40.00){\circle{2.00}}
\put(50.00,31.00){\vector(0,1){8.00}}
\put(50.00,50.00){\circle{2.00}}
\put(50.00,41.00){\vector(0,1){8.00}}
\put(50.00,60.00){\circle{2.00}}
\put(50.00,51.00){\vector(0,1){8.00}}
\put(50.00,70.00){\circle{2.00}}
\put(50.00,61.00){\vector(0,1){8.00}}
\put(50.00,80.00){\circle{2.00}}
\put(50.00,71.00){\vector(0,1){8.00}}
\put(50.00,90.00){\circle{2.00}}
\put(50.00,81.00){\vector(0,1){8.00}}
\put(80.00,10.00){\circle*{2.00}}
\put(90.00,10.00){\circle*{2.00}}
\put(100.00,10.00){\circle*{2.00}}
\put(110.00,10.00){\circle*{2.00}}
\put(80.00,20.00){\circle{2.00}}
\put(80.00,11.00){\vector(0,1){8.00}}
\put(80.00,30.00){\circle{2.00}}
\put(80.00,21.00){\vector(0,1){8.00}}
\put(90.00,20.00){\circle{2.00}}
\put(90.00,11.00){\vector(0,1){8.00}}
\put(90.00,30.00){\circle{2.00}}
\put(90.00,21.00){\vector(0,1){8.00}}
\put(100.00,20.00){\circle{2.00}}
\put(100.00,11.00){\vector(0,1){8.00}}
\put(100.00,30.00){\circle{2.00}}
\put(100.00,21.00){\vector(0,1){8.00}}
\put(100.00,40.00){\circle{2.00}}
\put(100.00,31.00){\vector(0,1){8.00}}
\put(100.00,50.00){\circle{2.00}}
\put(100.00,41.00){\vector(0,1){8.00}}
\put(100.00,60.00){\circle{2.00}}
\put(100.00,51.00){\vector(0,1){8.00}}
\put(100.00,70.00){\circle{2.00}}
\put(100.00,61.00){\vector(0,1){8.00}}
\put(100.00,80.00){\circle{2.00}}
\put(100.00,71.00){\vector(0,1){8.00}}
\put(100.00,90.00){\circle{2.00}}
\put(100.00,81.00){\vector(0,1){8.00}}
\put(110.00,20.00){\circle{2.00}}
\put(110.00,11.00){\vector(0,1){8.00}}
\put(110.00,30.00){\circle{2.00}}
\put(110.00,21.00){\vector(0,1){8.00}}
\put(110.00,40.00){\circle{2.00}}
\put(110.00,31.00){\vector(0,1){8.00}}
\put(110.00,50.00){\circle{2.00}}
\put(110.00,41.00){\vector(0,1){8.00}}
\put(120.00,10.00){\circle*{2.00}}
\put(120.00,20.00){\circle{2.00}}
\put(120.00,11.00){\vector(0,1){8.00}}
\put(120.00,30.00){\circle{2.00}}
\put(120.00,21.00){\vector(0,1){8.00}}
\put(120.00,40.00){\circle{2.00}}
\put(120.00,31.00){\vector(0,1){8.00}}
\put(120.00,50.00){\circle{2.00}}
\put(120.00,41.00){\vector(0,1){8.00}}
\put(80.00,31.00){\vector(2,1){18.67}}
\put(90.00,31.00){\vector(1,1){8.67}}
\put(110.00,51.00){\vector(-1,1){8.67}}
\put(119.67,51.00){\vector(-2,1){18.00}}
\put(10.00,5.00){\makebox(0,0)[cc]{$p_1$}}
\put(20.00,5.00){\makebox(0,0)[cc]{$p_2$}}
\put(30.00,5.00){\makebox(0,0)[cc]{$p_3$}}
\put(40.00,5.00){\makebox(0,0)[cc]{$p_4$}}
\put(50.00,5.00){\makebox(0,0)[cc]{$p_5$}}
\put(80.00,5.00){\makebox(0,0)[cc]{$p_1$}}
\put(90.00,5.00){\makebox(0,0)[cc]{$p_2$}}
\put(100.00,5.00){\makebox(0,0)[cc]{$p_3$}}
\put(110.00,5.00){\makebox(0,0)[cc]{$p_4$}}
\put(120.00,5.00){\makebox(0,0)[cc]{$p_5$}}
\put(5.00,95.00){\makebox(0,0)[cc]{a)}}
\put(75.00,95.00){\makebox(0,0)[cc]{b)}}
\put(145.00,95.00){\makebox(0,0)[cc]{c)}}
\put(170.00,90.00){\circle{2.00}}
\put(170.00,80.00){\circle{2.00}}
\put(170.00,70.00){\circle{2.00}}
\put(170.00,60.00){\circle{2.00}}
\put(170.00,50.00){\circle{2.00}}
\put(170.00,40.00){\circle{2.00}}
\put(170.00,30.00){\circle{2.00}}
\put(170.00,20.00){\circle{2.00}}
\put(170.00,10.00){\circle*{2.00}}
\put(170.00,5.00){\makebox(0,0)[cc]{$p_1$}}
\put(170.00,11.00){\vector(0,1){8.00}}
\put(170.00,21.00){\vector(0,1){8.00}}
\put(170.00,31.00){\vector(0,1){8.00}}
\put(170.00,41.00){\vector(0,1){8.00}}
\put(170.00,51.00){\vector(0,1){8.00}}
\put(170.00,61.00){\vector(0,1){8.00}}
\put(170.00,71.00){\vector(0,1){8.00}}
\put(170.00,81.00){\vector(0,1){8.00}}
\put(280.00,131.00){\vector(0,1){8.00}}
\put(280.00,141.00){\vector(0,1){8.00}}
\put(180.00,70.00){\circle{2.00}}
\put(190.00,70.00){\circle{2.00}}
\put(180.00,80.00){\circle{2.00}}
\put(180.00,71.00){\vector(0,1){8.00}}
\put(180.00,90.00){\circle{2.00}}
\put(180.00,81.00){\vector(0,1){8.00}}
\put(190.00,80.00){\circle{2.00}}
\put(190.00,71.00){\vector(0,1){8.00}}
\put(190.00,90.00){\circle{2.00}}
\put(190.00,81.00){\vector(0,1){8.00}}
\put(150.00,50.00){\circle{2.00}}
\put(150.00,60.00){\circle{2.00}}
\put(150.00,51.00){\vector(0,1){8.00}}
\put(150.00,70.00){\circle{2.00}}
\put(150.00,61.00){\vector(0,1){8.00}}
\put(150.00,80.00){\circle{2.00}}
\put(150.00,71.00){\vector(0,1){8.00}}
\put(150.00,90.00){\circle{2.00}}
\put(150.00,81.00){\vector(0,1){8.00}}
\put(160.00,50.00){\circle{2.00}}
\put(160.00,60.00){\circle{2.00}}
\put(160.00,51.00){\vector(0,1){8.00}}
\put(160.00,70.00){\circle{2.00}}
\put(160.00,61.00){\vector(0,1){8.00}}
\put(160.00,80.00){\circle{2.00}}
\put(160.00,71.00){\vector(0,1){8.00}}
\put(160.00,90.00){\circle{2.00}}
\put(160.00,81.00){\vector(0,1){8.00}}
\put(168.67,39.67){\vector(-2,1){18.00}}
\put(169.00,40.33){\vector(-1,1){8.33}}
\put(170.67,60.33){\vector(1,1){8.67}}
\put(170.67,59.67){\vector(2,1){18.67}}
\end{picture}
\end{center}
\caption{The lowest ``root'' represents the
initial state of the computer. Forward computation represents
upwards motion
through a sequence of states represented by open circles. Different
symbols $p_i$ correspond to different initial computer states.
a) One-to-one computation.
b) Many-to-one junction which is information discarding. Several
computational paths, moving upwards, merge into one.
c) One-to-many computation is allowed only
 if no information is
created and discarded; e.g., in copy-type operations on blank memory.
From Landauer \protect\cite{landauer-94}.
\label{f-rev-comp}
}
\end{figure}
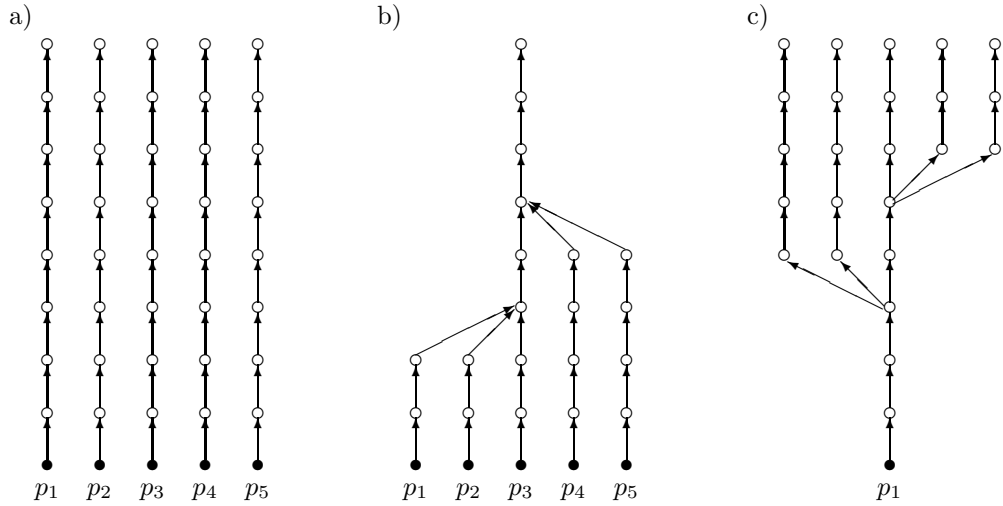

The connection between information and physical entropy, in particular
the entropy increase during computational steps corresponding to an
irreversible loss of information---deletion or other many-to-one
operations---has raised considerable attention in the physics community
\cite{maxwell-demon}.
Figure \ref{f-rev-comp}  \cite{landauer-94} draws the difference
between one-to-one, many-to-one and one-to-many computation.
Classical
reversible computation
\cite{landauer:61,bennett-73,fred-tof-82,bennett-82,landauer-94}
is characterized by a single-valued inverse
evolution function.
In such cases
it is always possible to ``reverse the gear'' of the evolution, and
compute the input from the output, the initial state from the final
state.

In irreversible computations, logical functions
are performed which
do not have a single-valued inverse, such as ${and}$ or ${or}$;
i.e., the input cannot be deduced from the output. Also deletion of
information or other many
(states)-to-one
(state) operations are irreversible.
This logical irreversibility is associated with physical irreversibility
and requires a minimal heat generation of the computing machine and
thus an entropy increase.

It is possible to embed any irreversible computation in an appropriate
environment which makes it reversible. For instance, the computer
could keep the inputs of previous calculations in successive order.
It could save all the information it would otherwise throw away.
 Or,
it could leave markers behind to identify its trail, the {\it H\"ansel
and Gretel} strategy described by Landauer \cite{landauer-94}. That, of
course, might amount to huge overheads in dynamical memory space
(and time) and would merely postpone the problem of throwing away
unwanted information. But, as has been pointed out by Bennett
\cite{bennett-73}, for classical computations, in which copying and
one-to-many operations are still allowed, this overhead could be
circumvented by
erasing all intermediate results, leaving behind only copies of the
output and the original input. Bennett's trick is
to perform  a computation,  copy the resulting output
and then, with one output as input, run
the computation backward. In order not to consume exceedingly large
intermediate storage resources, this strategy could be applied after
every single step.
Notice that copying can be done
reversible in classical physics if the memory used for the copy is
initially considered to be blank.

Quantum mechanics, in particular quantum computing, teaches us to
restrict ourselves even more and exclude any one-to-many operations, in
particular copying, and to accept merely one-to-one
computational operations
corresponding to bijective mappings [cf.
Figure \ref{f-rev-comp}a)].
This is due to the fact that the unitary
evolution of the quantum mechanical state state (between two subsequent
measurements) is strictly one-to-one.
Per definition, the inverse of a unitary operator $U$ representing a
quantum mechanical time evolution always exists. It is again a unitary
operator $U^{-1}=U^\dagger$ (where $\dagger$ represents the adjoint
operator); i.e., $UU^\dagger =1$.

The {\em no-cloning theorem}
\cite{herbert,wo-zu,mandel:83,mil-hard,glauber,caves}
prevents one-to-many
operations, in particular the copying of general (nonclassical)
quantum bits of information. Thus Bennett's original strategy cannot be
applied in the case one-to-one computations or quantum computations.

In what follows we shall consider a particular example of a
one-to-one deterministic computation.  Although tentative in its present
form, this example may illustrate the conceptual strength of reversible
computation.  Our starting point are
finite automata \cite{e-f-moore,conway,brauer-84,schaller-96,cal-sv-yu},
but of a very particular,
hithero unknown sort.  They are characterized by a finite set $S$ of
states, a finite input and output alphabet $I$ and $O$, respectively.
Like for Mealy automata, their temporal evolution and output functions
are given by $\delta :S\times I\rightarrow S$, $\lambda :S\times
I\rightarrow O$.  We additionally require one-to-one reversibility,
which we interpret in this context as follows.  Let $I=O$, and let the
combined (state and output) temporal evolution be associated with a
bijective map
\begin{equation}
U:(s,i)\rightarrow (\delta(s,i),\lambda (s,i)),
\label{t-e-l}
\end{equation}
with
$s\in S$ and $i\in I$.
The state and output symbol could be ``fed back'' consecutively; such
that
$N$ evolution steps correspond to $U^N=\underbrace{U\cdots U}_{N
\;\textrm{times}}$.

The elements of the Cartesian product
$S\times I$ can be arranged as a linear list of length
$n$ corresponding to  a vector. In this sense,
$U$ corresponds to a $n\times n$-matrix.  Let $\Psi_i$
be the $i$'th element in the vectorial representation of some
$(s,i)$, and let
$U_{ij}$ be the element
of
$U$ in the $i$'th row and the $j$'th column.
Due to
determinism, uniqueness and invertibility,
\begin{description}
\item[$\bullet$]
$U_{ij}\in \{0,1\}$;
\item[$\bullet$]
orthogonality:
 $U^{-1}=U^t$ (superscript $t$ means transposition) and
$(U^{-1})_{ij}=U_{ji}$;
\item[$\bullet$]
double stochasticity:
the sum of each row and column is one; i.e.,
$\sum_{i=1}^n U_{ij}= \sum_{j=1}^n U_{ij}=1$.
\end{description}
Since $U$ is a square matrix whose elements are either one or zero and
which has exactly one nonzero entry in each row and exactly one in each
column, it is a {\em permutation matrix}.
Let ${\cal P}_n$ denote the set of all $n\times n$ permutation matrices.
 ${\cal P}_n$ forms the {\em Permutation group} (sometimes called {\em
symmetric group}) of degree $n$. (The product of two permutation
matrices is
a permutation matrix, the inverse is the transpose and the identity
${\bf
1}$ belongs to  ${\cal P}_n$.)
${\cal P}_n$ has $n!$ elements.
Furthermore, the set of all doubly stochastic matrices forms a convex
polyhedron with the permutation matrices as vertices
\cite[page 82]{ber-ple}.

Let us be more specific.
For $n=1$, ${\cal P}_1=\{1\}$.\\
For $n=2$, ${\cal P}_2=\left\{
\left(
\begin{array}{cc}
1&0\\
0&1
\end{array}
\right)
,
\left(
\begin{array}{cc}
0&1\\
1&0
\end{array}
\right)
\right\}$.\\
For $n=3$,
${\cal P}_3=\left\{
\left(
\begin{array}{ccc}
1&0&0\\
0&1&0\\
0&0&1
\end{array}
\right)
,
\left(
\begin{array}{ccc}
1&0&0\\
0&0&1\\
0&1&0
\end{array}
\right)
,
\left(
\begin{array}{ccc}
0&1&0\\
0&0&1\\
1&0&0
\end{array}
\right)
,
\left(
\begin{array}{ccc}
0&1&0\\
1&0&0\\
0&0&1
\end{array}
\right)
,
\left(
\begin{array}{ccc}
0&0&1\\
1&0&0\\
0&1&0
\end{array}
\right)
,
\left(
\begin{array}{ccc}
0&0&1\\
0&1&0\\
1&0&0
\end{array}
\right)\right\}
.           $

The correspondence between permutation matrices and
reversible automata is straightforward. Per definition [cf. Equation
(\ref{t-e-l})], every reversible automaton is representable by some
permutation matrix. That every $n\times n$ permutation matrix
corresponds to an automaton can be demonstrated by
considering the simplest
case of a one state automaton with $n$ input/output symbols.
There exist less trivial identifications. For example,
let $$\widetilde{ U}=
\left(
\begin{array}{cccc}
0&0&1&0\\
0&1&0&0\\
0&0&0&1\\
1&0&0&0
\end{array}
\right).
$$
The transition and output functions of one associated reversible
automaton is listed in table
\ref{t-ra}.
\begin{table}
\begin{center}
\begin{tabular}{|c|cc|cc|}
 \hline
 \hline
 &$\delta$ & & $\lambda$&\\
$S\backslash I$ &1&2& 1&2\\
 \hline
$s_1$&$s_2$&$s_1$ & 1&2\\
$s_2$&$s_2$&$s_1 $& 2&1\\
 \hline
 \hline
\end{tabular}
\end{center}
\caption{Transition and output table of a reversible
automaton with two states $S=\{s_1, s_2\}$ and two input/output
symbols $I= \{1,2\}$.\label{t-ra}}
\end{table}
Since $\widetilde{U}$ has a cycle 3; i.e., $(\widetilde{U})^3={\bf 1}$,
irrespective of the initial state, the automaton is back at its initial
state after three evolution steps. For example,
$
(s_1,1)\rightarrow
(s_2,1)\rightarrow
(s_2,2)\rightarrow
(s_1,1)$.

The discrete temporal evolution (\ref{t-e-l}) can, in matrix
notation, be represented by
\begin{equation}
U\Psi (N)= \Psi (N+1)=U^{N+1}\Psi (0),
\label{t-e-l2}
\end{equation}
where again $N=0,1,2,3,\ldots$ is a discrete time parameter.



\begin{figure}
\begin{center}
\unitlength 0.9mm
\linethickness{0.4pt}
\begin{picture}(79.33,115.00)
\put(2.33,14.67){\framebox(5.00,5.00)[cc]{$\footnotesize i_{1}$}}
\put(9.33,14.67){\framebox(5.00,5.00)[cc]{$\footnotesize i_{2}$}}
\put(16.33,14.67){\framebox(5.00,5.00)[cc]{$\footnotesize i_{3}$}}
\put(23.33,14.67){\framebox(5.00,5.00)[cc]{$\footnotesize i_{4}$}}
\put(30.33,14.67){\framebox(5.00,5.00)[cc]{$\footnotesize i_{5}$}}
\put(37.33,14.67){\framebox(5.00,5.00)[cc]{$\footnotesize i_{6}$}}
\put(44.33,14.67){\framebox(5.00,5.00)[cc]{$\footnotesize i_{7}$}}
\put(51.33,14.67){\framebox(5.00,5.00)[cc]{$\footnotesize i_{8}$}}
\put(58.33,14.67){\framebox(5.00,5.00)[cc]{$\footnotesize i_{9}$}}
\put(65.33,14.67){\framebox(5.00,5.00)[cc]{$\footnotesize i_{10}$}}
\put(72.33,14.67){\framebox(5.00,5.00)[cc]{$\footnotesize i_{11}$}}
\put(2.33,7.67){\framebox(5.00,5.00)[cc]{$\footnotesize i_{12}$}}
\put(9.33,7.67){\framebox(5.00,5.00)[cc]{$\footnotesize i_{13}$}}
\put(16.33,7.67){\framebox(5.00,5.00)[cc]{$\footnotesize i_{14}$}}
\put(23.33,7.67){\framebox(5.00,5.00)[cc]{$\footnotesize i_{15}$}}
\put(30.33,7.67){\framebox(5.00,5.00)[cc]{$\footnotesize i_{16}$}}
\put(37.33,7.67){\framebox(5.00,5.00)[cc]{$\footnotesize i_{17}$}}
\put(44.33,7.67){\framebox(5.00,5.00)[cc]{$\footnotesize i_{18}$}}
\put(51.33,7.67){\framebox(5.00,5.00)[cc]{$\footnotesize i_{19}$}}
\put(58.33,7.67){\framebox(5.00,5.00)[cc]{$\cdot$}}
\put(65.33,7.67){\framebox(5.00,5.00)[cc]{$\cdot$}}
\put(72.33,7.67){\framebox(5.00,5.00)[cc]{$\cdot$}}
\put(20.00,35.00){\line(1,0){30.00}}
\put(50.00,35.00){\line(0,1){25.00}}
\put(50.00,60.00){\line(-1,0){30.00}}
\put(20.00,60.00){\line(0,-1){25.00}}
\put(50.00,60.00){\line(3,2){15.00}}
\put(20.00,60.00){\line(3,2){15.00}}
\put(50.00,35.00){\line(3,2){15.00}}
\put(65.00,45.00){\line(0,1){25.00}}
\put(65.00,70.00){\line(-1,0){30.00}}
\put(0.00,5.00){\line(1,0){79.33}}
\put(79.33,5.00){\line(0,1){17.00}}
\put(79.33,22.00){\line(-1,0){79.33}}
\put(0.00,22.00){\line(0,-1){17.00}}
\put(30.00,90.00){\line(1,0){20.00}}
\put(50.00,90.00){\line(0,1){20.00}}
\put(50.00,110.00){\line(-1,0){20.00}}
\put(30.00,110.00){\line(0,-1){20.00}}
\multiput(40.67,66.33)(0.11,0.16){15}{\line(0,1){0.16}}
\multiput(42.35,68.80)(0.12,0.20){11}{\line(0,1){0.20}}
\multiput(43.67,70.99)(0.12,0.24){8}{\line(0,1){0.24}}
\multiput(44.62,72.90)(0.12,0.33){5}{\line(0,1){0.33}}
\multiput(45.20,74.55)(0.11,0.69){2}{\line(0,1){0.69}}
\multiput(45.42,75.92)(-0.08,0.55){2}{\line(0,1){0.55}}
\multiput(45.27,77.01)(-0.10,0.16){5}{\line(0,1){0.16}}
\multiput(44.75,77.84)(-0.18,0.11){5}{\line(-1,0){0.18}}
\multiput(43.87,78.39)(-0.42,0.09){3}{\line(-1,0){0.42}}
\put(42.62,78.66){\line(-1,0){1.62}}
\put(41.00,78.67){\line(-1,0){1.85}}
\multiput(39.15,78.71)(-0.50,0.08){3}{\line(-1,0){0.50}}
\multiput(37.66,78.95)(-0.28,0.11){4}{\line(-1,0){0.28}}
\multiput(36.53,79.41)(-0.13,0.11){6}{\line(-1,0){0.13}}
\multiput(35.78,80.07)(-0.10,0.22){4}{\line(0,1){0.22}}
\put(35.39,80.93){\line(0,1){1.07}}
\multiput(35.36,82.00)(0.11,0.43){3}{\line(0,1){0.43}}
\multiput(35.70,83.28)(0.12,0.25){6}{\line(0,1){0.25}}
\multiput(36.41,84.76)(0.12,0.19){9}{\line(0,1){0.19}}
\multiput(37.48,86.44)(0.12,0.15){24}{\line(0,1){0.15}}
\multiput(39.67,34.67)(0.18,-0.11){9}{\line(1,0){0.18}}
\multiput(41.31,33.67)(0.11,-0.13){8}{\line(0,-1){0.13}}
\put(42.18,32.59){\line(0,-1){1.16}}
\multiput(42.29,31.43)(-0.11,-0.21){6}{\line(0,-1){0.21}}
\multiput(41.64,30.18)(-0.14,-0.11){19}{\line(-1,0){0.14}}
\multiput(39.00,28.00)(-0.22,-0.12){8}{\line(-1,0){0.22}}
\multiput(37.23,27.05)(-0.12,-0.13){8}{\line(0,-1){0.13}}
\multiput(36.27,26.01)(-0.07,-0.57){2}{\line(0,-1){0.57}}
\multiput(36.12,24.86)(0.11,-0.21){6}{\line(0,-1){0.21}}
\multiput(36.79,23.62)(0.13,-0.12){14}{\line(1,0){0.13}}
\put(35.00,47.33){\makebox(0,0)[cc]{\huge ?}}
\put(40.00,100.00){\makebox(0,0)[cc]{$o_j$}}
\put(40.00,0.00){\makebox(0,0)[cc]{input interface}}
\put(40.00,115.00){\makebox(0,0)[cc]{output interface}}
\put(70.00,55.00){\makebox(0,0)[lc]{box}}
\end{picture}
\end{center}
\caption{\label{f-bbox}
A system in a black box
with an input interface and an output interface.}
\end{figure}
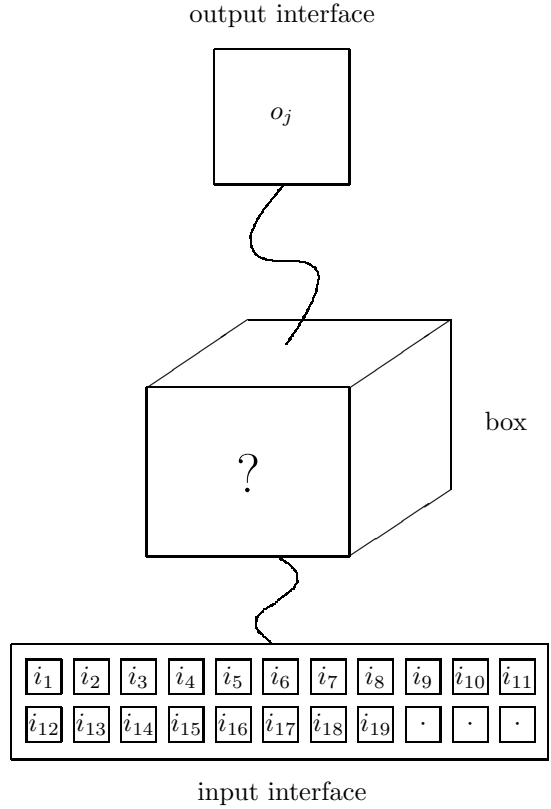

Let us come back to our original issue of modelling the measurement
process within a system whose states evolve according  to a one-to-one
evolution.
Let us artificially divide such a system into an ``inside'' and an
``outside'' region.
This can be suitably represented by introducing a black box which
contains the ``inside'' region---the subsystem to be measured, whereas
the
remaining ``outside'' region is interpreted as the measurement
apparatus.
An input and an output interface mediate all interactions of the
``inside'' with the ``outside,'' of the ``observed'' and the
``observer'' by symbolic exchange. Let us assume that, despite such
symbolic exchanges via the interfaces
(for all practical purposes),
to an outside
observer
 what happens inside the black box is a hidden,
 inaccessible
  arena.
The observed system is like the ``black box'' drawn in
Figure~\ref{f-bbox}.

Throughout temporal evolution, not only is information transformed
one-to-one
(bijectively, homomorphically) inside the black box,
but
this information is handled  one-to-one {\em after} it appeared on the
black box
interfaces. It might  seem evident at first glance that the symbols
appearing on the interfaces should be treated as classical information.
That is, they could in principle be copied. The possibility to copy the
experiment (input and output) enables the application of Bennett's
argument: in such a case, one keeps the experimental finding by copying
it, revert the system evolution and starts with a ``fresh'' black
box system in its original initial state. The result is a
classical Boolean calculus.

The scenario is drastically changed, however, if we assume a
one-to-one evolution also for the environment at and outside
of the black box. That is, one deals with a
homogeneous
and uniform one-to-one evolution ``inside'' and ``outside'' of the black
box, thereby
assuming that the experimenter also evolves
one-to-one and not classically.
In our toy automaton model, this could for instance be realized by some
automaton corresponding to a permutation operator $U$ inside the black
box, and another reversible automaton corresponding to another $U'$
outside of it. Conventionally, $U$ and $U'$ correspond to the measured
system and the measurement device, respectively.

In such a case, as there is no copying due to one-to-one evolution,
in order to set
back the system to its original initial state, the experimenter would
have to erase all knowledge bits of information acquired so far.
The experiment would have to evolve back to the initial state of the
measurement device and the measured system prior to the measurement.
As a result, the representation of measurement results in one-to-one
reversible systems may cause a sort of complementarity due to
the impossibility to measure all variants of the representation
at once.

Let us give a brief example. Consider the $6\times 6$ permutation matrix
$$
U=
\left(
\begin{array}{cccccc}
0&1&0&0&0&0\\
0&0&0&0&0&1\\
0&0&1&0&0&0\\
1&0&0&0&0&0\\
0&0&0&0&1&0\\
0&0&0&1&0&0
\end{array}
\right)
$$
corresponding to a reversible $3$-state automaton with two input/output
symbols $1,2$.
listed in table
\ref{t-rra}.
\begin{table}
\begin{center}
\begin{tabular}{|c|cc|cc|}
 \hline
 \hline
 &$\delta$ & & $\lambda$&\\
$S\backslash I$ &1&2& 1&2\\
 \hline
$s_1$&$s_1$&$s_3$ & 2&2\\
$s_2$&$s_2$&$s_1 $& 1&1\\
$s_3$&$s_3$&$s_2 $& 1&2\\
 \hline
 \hline
\end{tabular}
\end{center}
\caption{Transition and output table of a reversible
automaton with three states $S=\{s_1, s_2, s_3\}$ and two input/output
symbols $I= \{1,2\}$.\label{t-rra}}
\end{table}
The evolution is
$$
\left(
\begin{array}{c}
(s_1,1)                \\
(s_1,2)                \\
(s_2,1)                \\
(s_2,2)                \\
(s_3,1)                \\
(s_3,2)
\end{array}
\right)
\stackrel{U}{\longrightarrow }
\left(
\begin{array}{c}
(s_1,2)                \\
(s_3,2)                \\
(s_2,1)                \\
(s_1,1)                \\
(s_3,1)                \\
(s_2,2)
\end{array}
\right)
\stackrel{U}{\longrightarrow }
\left(
\begin{array}{c}
(s_3,2)                \\
(s_2,2)                \\
(s_2,1)                \\
(s_1,2)                \\
(s_3,1)                \\
(s_1,1)
\end{array}
\right)
\stackrel{U}{\longrightarrow }
\left(
\begin{array}{c}
(s_2,2)                \\
(s_1,1)                \\
(s_2,1)                \\
(s_3,2)                \\
(s_3,1)                \\
(s_1,2)
\end{array}
\right)
\stackrel{U}{\longrightarrow }
\left(
\begin{array}{c}
(s_1,1)                \\
(s_1,2)                \\
(s_2,1)                \\
(s_2,2)                \\
(s_3,1)                \\
(s_3,2)
\end{array}
\right).
  $$
Thus after the input of just one symbol, the automaton states can be
grouped into experimental equivalence classes \cite{svozil-93}
$$v(1)=\{\{1\},\{2,3\}\},\quad
v(2)=\{\{1,3\},\{2\}\}.$$
The associated partition logic corresponds to a non Boolean
(nondistributive)
partition logic isomorphic to $MO_2$. Of course, if one develops the
automaton further, then, for instance, $v(2222)=\{\{1\},\{2\},\{3\}\}$
and the classical case is recovered [notice that this is not the case
for $v(\stackrel{\cdot}{1})=v(1)$]. Yet, if one assumes that the output
is channelled into the interface after only a single evolution step (and
that afterwards the evolution is via another $U'$), the
nonclassical feature pertains despite the bijective character of the
evolution.

What has been discussed above is very similar to the
opening,
closing and reopening of Schr\"odinger's catalogue of expectation values
\cite[p. 53]{schrodinger}:
At least up to
a certain magnitude of complexity---any measurement can be ``undone'' by
a proper reconstruction of the wave-function. A necessary condition for
this to happen is that {\em all} information about the original
measurement is lost.
In Schr\"odinger's terms,
the prediction catalog
(the wave function) can be opened only at one particular page.
We may close
the prediction catalog
before reading this page. Then we can open
the prediction catalog
at another, complementary, page again.
By no way we can open
the prediction catalog at one page, read and (irreversible) memorize the
page, close it; then open it at another, complementary, page.
(Two noncomplementary pages which correspond to two co-measurable
observables can be read simultaneously.)

From this point of view, it appears that, strictly speaking,
irreversibility
may turn out to be an inappropriate concept both in computational
universes generated by one-to-one evolution as well as for quantum
measurement theory. Indeed, irreversibility may have been imposed upon
the measurement process
rather heuristically and artificially to express the huge practical
difficulties associated with any
backward evolution, with ``reversing the gear'', or with reconstructing
a coherent state.
 To quote Landauer
\cite[section 2]{landauer-89},
{\em ``What is measurement? If it is simply information transfer, that
is done all the time inside the computer, and can be done with arbitrary
little dissipation.''} And, one may add, {\em without destroying
coherence.}

Let us conclude with a metaphysical speculation.
In a one-to-one invertible universe, any  evolution, any step of
computation, any single measurement
act reminds us of a permanent permutation, reformulation and reiteration
of one
and the same ``message''---a ``message'' that was there already at
beginning of the universe, which gets transformed but is neither
destroyed nor renewed. This thought might be very close to what
Schr\"odinger had in mind when contemplating about Vedic philosophy
\cite{schroed:welt}.


\end{document}